\documentclass[a4paper,12pt]{article}
\usepackage{graphicx}
\DeclareGraphicsExtensions{.jpg}

\begin{document}

\begin{center}
\begin{Large}
\textbf{Underwater Gas Expansion and Deflagration}\\
\

\end{Large}

\
\

\includegraphics[height=60mm]{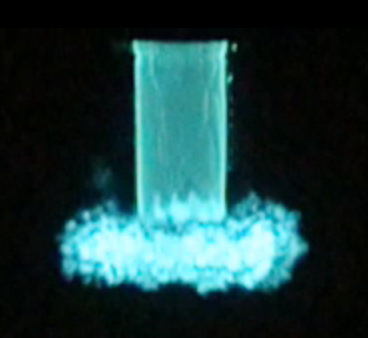}
\includegraphics[height=60mm]{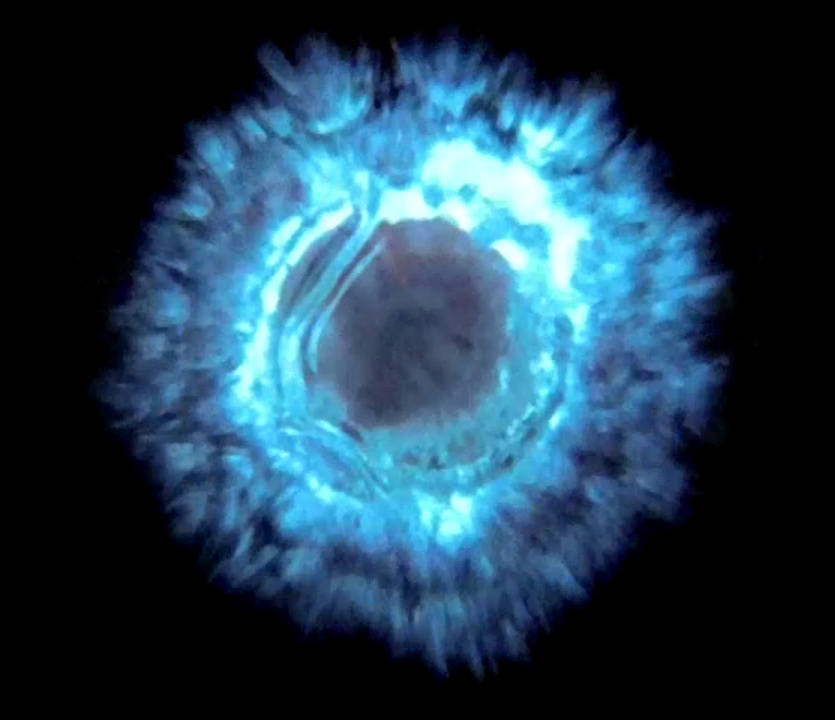}\\

\

\

Van Jones, Kariann Vander Pol, John Gilbert, Leigh McCue-Weil

\

Department of Aerospace and Ocean Engineering,\\
Virginia Tech, Blacksburg, VA 24060, USA \\
October 5, 2013

\

\

\textbf{Abstract} \\
The underwater combustion of a propane-air mixture in an acrylic cylinder is captured on video from multiple angles. This experiment is designed to provide visual data and pressure time-histories for future CFD validation studies.

\

\end{center}

\begin{flushleft}
\textbf{Introduction:}\\
\end{flushleft}

Smoothed Particle Hydrodynamics (SPH) is a Lagrangian CFD method based on approximating point numerical field values from a finite integral of the surrounding fluid field [1]. For near-boundary fluid points this volume integral can become truncated. This breaks the assumption that well defined field data surrounds any given point and can lead to erroneous calculations of field values in boundary regions. Because of this, SPH fluid simulations often require additional methods to correct for or eliminate the this truncation error in near-boundary regions.

In order to aid in the validation of an SPH code developed by the authors; a series of experiments has been designed to provide visual and pressure time-history data to compare with simulation results. The experimental apparatus consists of a frame immersed in water. Mounted to the frame is an inverted cylinder filled with a flammable propane-air mixture (Figure 1). A traversable plate embedded with a row of pressure sensors is to be mounted beneath the cylinder (not shown). 

The video submission is a compilation of initial test runs performed to analyze the combustion of the fuel-air mixture. The primary goal of the initial tests runs was to determine an ideal set of fuel-air mixtures and volumes for later testing. A secondary interest of the tests is the analysis of the flame-front propagation through the fuel-air mixture. This is to provide insight into the homogeneity of the fuel-air mixture and to verify that modeling the combustion even using a point-ignition source is accurate.

\
\
\

\begin{center}
\includegraphics[height=100mm]{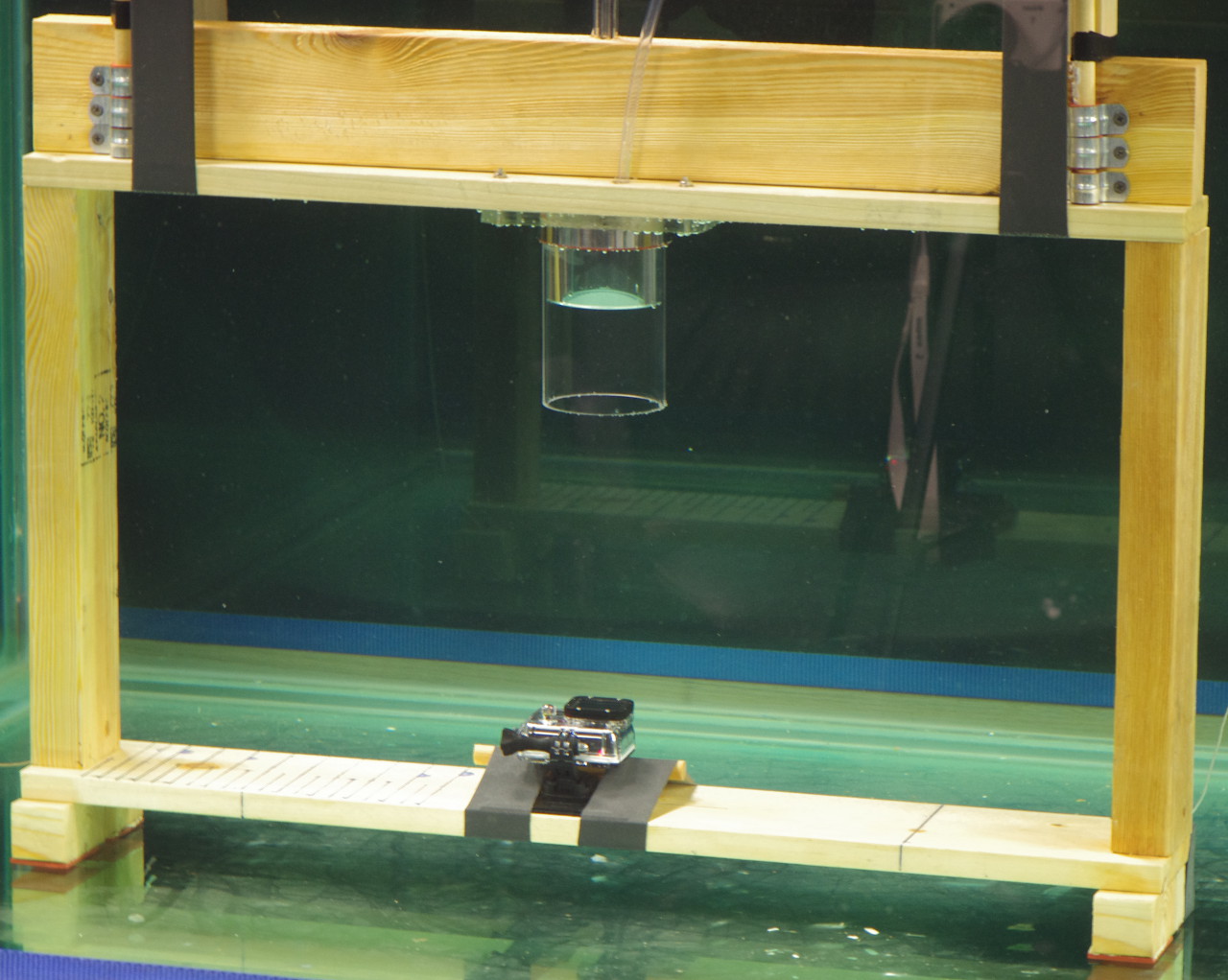}\\
Figure 1: Experiment instrumented with upward-looking camera
\end{center}

\
\
\
\

\

\begin{flushleft}
\

\textbf{Video Description:}
\end{flushleft}

An inverted acrylic cylinder is immersed in water and filled with a propane-air mixture. Ignition is initiated by a spark that arcs across two electrodes located top-center of the cylinder. This fluid dynamics video is a compilation of recordings taken from several angles for analysis of the resultant explosion and fluid motion. 

The video shows the progress of the explosion event. Once the fuel-air mixture is ignited, a flame-front propagates downward through the fuel-air mixture. During the combustion process, high pressure gas is expelled from the cylinder and expands to form a non-spherical gas bubble. After combustion is complete, the overexpanded gas-bubble collapses, forming a jet pair aligned with the axis of the cylinder. The upward jet extends into the cylinder interior before impinging on the cylinder’s upper surface. The downward jet travels into the fluid and is made visible by captured gas bubbles convected along with the jet.

A bottom-view of a smaller combustion event is shown with and without external illumination. This highlights the spread of the flame-front from the electrodes and the subsequent burning of fuel in the mixture through the bubble expansion phase.

\
\
\

\begin{flushleft}
\textbf{Specifications:}
\end{flushleft}

\begin{itemize}

\item[] Cylinder:
\begin{itemize}
\item[]		Diameter:   3 inches\
\item[]     Length: ~5 inches\
\end{itemize}

\item[] Tank:
\begin{itemize}
\item[]		4’x4’x4’  (~500 gallons)
\end{itemize}

\item[] Air-fuel Mixture:
\begin{itemize}
\item[]		6\% Propane by volume
\item[]		Total mixture volume: 300 - 600 ml
\end{itemize}

\item[] Cameras:
\begin{itemize}
	\item[]	Nikon J1 (external footage):
		\begin{itemize}
		\item[]	400fps (640x240)
		\item[] 1200 fps - (320x120)
		\end{itemize}
	
	\item[] GoPro Hero 3 (underwater footage): 
		\begin{itemize}
		\item[]	240 fps (848x480)
		\end{itemize}
\end{itemize}

\begin{flushleft}
\textbf{Acknowledgements:}\
\end{flushleft}

This research was made possible by ONR grant N000141010398 with special thanks to Pat Purtell and Ki-Han Kim.

\begin{flushleft}
\textbf{Related Papers:}\
\end{flushleft}

\item[]
[1] Gingold RA, Monaghan JJ. "Smoothed particle hydrodynamics-theory and
application to non-spherical stars". \textit{Mon Notic Roy Astron Soc.}
1977;181:375–89.

\item[]
Jones, V., Yang, Q., and McCue, L., "SPH Boundary Deficiency Correction for Improved Boundary Conditions at Deformable Surfaces," \textit{Ship Science and Technology/Ciencia y Tecnología de Buques}, Volume 4, Number 7, July 2010.

\item[]
Yang, Q., Jones, V., and McCue, L., "Investigation of skirt dynamics of air cushion vehicles under non-linear wave impact using a SPH-FEM model," \textit{11th International Conference on Fast Sea Transportation (FAST)}, Honolulu, HI, September, 2011.

\item[]
Yang, Q., Jones, V., and McCue, L., "Free-surface Flow Interactions with Deformable Structures Using an SPH-FEM Model," \textit{Ocean Engineering}, Volume 55, December, 2012, pp. 136-147.
\end{itemize}

\end{document}